\documentclass{revtex4-2}

\usepackage[utf8]{inputenc}
\DeclareUnicodeCharacter{2264}{$\leq$}
\usepackage{graphicx}

\begin{document}

\title{{Impact of spin correlations on resistivity and microwave absorption of Ba(Fe$_{1-x}$Co$_x$)$_2$As$_2$}}

\author{Yu.~I.~Talanov} \email{talanov@kfti.knc.ru}
\affiliation{Zavoisky Physical-Technical Institute\\ 420029, Kazan, Russia}

\author{I.~I.~Gimazov} 
\affiliation{Zavoisky Physical-Technical Institute\\ 420029, Kazan, Russia}

\author{D.~E.~Zhelezniakova} 
\affiliation{Zavoisky Physical-Technical Institute\\ 420029, Kazan, Russia}

\begin{abstract}
The results of studies of BaFe$_2$As$_2$ single crystals doped with cobalt by means of resistivity and microwave absorption measurement are reported. A theoretical description of the behavior of the microwave absorption amplitude is made taking into account the temperature dependence of resistivity, magnetic susceptibility and the lifetime of spin fluctuations. An assumption has been made that the deviation from the linear dependence of resistivity on temperature at $T<100$\,K is not related to the electron-electron scattering mechanism, but it is due to the appearance of nematic fluctuations. Estimates of the rate of scattering by spin fluctuations indicate their nematic nature at temperatures near the structural transition. 
\end{abstract}

\maketitle

\section{Introduction}

Spin correlations are mostly considered as the main candidate for the role of mediator between free electrons when they combine into Cooper pairs upon emerging a superconducting state in iron based superconductors, in particular in the iron arsenide compounds. An essential reason for this statement is the fact that they determine the behavior of many physical parameters, in particular, the temperature dependence of resistance in the range above the superconducting transition, $T>T_c$\ (see, f.e.\cite{Ni2008,Wang2009,Doiron2009,Nakajima2014}). In several theoretical papers \cite{Moriya1990,Hlubina1995,Kemper2011,Fernandes2011}, it was shown that the experimentally observed linear dependence of resistivity on temperature $\rho(T)\propto T$\ (instead of the quadratic one characteristic of the scattering of quasiparticles in a Fermi liquid $\rho(T)\propto T^2$) is a consequence of their scattering just by spin fluctuations. Since antiferromagnetic spin correlations have a highly anisotropic order parameter corresponding to ordering in the form of spin density waves (SDW), they stimulate the establishment of nematic order in the temperature region near the transition point to the SDW phase $T_N$\ and nematic fluctuations above it (see Ref.\cite{Boehmer2022} and references there). The latter manifest themselves both in magnetic measurements and in electronic conductivity in a wide temperature range, which extends to tens of degrees Kelvin above $T_N$. There is a statement that nematic fluctuations serve as a trigger for a structural transition from the symmetrical tetragonal crystal structure of C$_4$\ to the anisotropic orthorhombic one with the C$_2$\ symmetry \cite{Nandi2010,Fernandes2010}. This transition occurs with a decrease in temperature at $T=T_s$, which is several degrees higher than the magnetic transition $T_N$.

Moreover, some authors point out the connection between transport properties (in particular, the contribution of spin fluctuations to the rate of carrier scattering) and superconducting parameters, such as the critical temperature $T_c$\ and the symmetry of the order parameter \cite{Doiron2009}. When studying the influence of different contributions to the scattering process, it is important to estimate and compare the scattering rates at centers of different types: defects ($\tau^{-1}_d$), impurities ($\tau^{-1}_{imp}$), phonons ($\tau^{-1}_{ph}$), spin fluctuations ($\tau^{-1}_{sf}$), \emph{etc}. The value of the total rate ($\tau^{-1}_{total}$) can be determined using the results of measuring the resistance of the material to direct current. And to evaluate the spin-fluctuation contribution, it is necessary to study the response to an alternating electromagnetic field, which changes with a frequency $\omega$\ comparable to the rate of scattering by spin fluctuations $\tau^{-1}_{sf}$. For this purpose, in this work we used the method of recording microwave absorption at a frequency of $\sim 10^{10}$\,Hz.

\section{Experimental methods and technique}

According to the classical Drude's theory, the resistivity of a material with metallic conductivity is determined by a small set of its physical parameters: 
\begin{equation}
\rho = \frac{m^*}{e^2n}\tau^{-1}
\label{eq1}
\end{equation}
where $m^*$, $e$ and $n$ are an effective mass, charge and concentration of current carriers (electrons or holes). The scattering rates $\tau^{-1}$\ consists of several contributions (as noted above): (i) scattering by defects and impurities $\tau^{-1}_d$; (ii) an "electron-electron scattering" of Fermi liquid quasiparticles upon interaction (collision) with each other $\tau^{-1}_e$; (iii) scattering by magnetic (spin) fluctuations $\tau^{-1}_{sf}$, if the material has magnetic moments interacting with each other. The presence of other scattering processes is also possible, but in the compounds we studied, only those listed above act. Since the dependence of resistivity on temperature is determined mainly by changes in scattering rates, it can be described by the sum of three terms: $\rho(T)=\rho_o+\rho_e(T)+\rho_{sf}(T)$. The first term $\rho_o$, which due to scattering by impurities and defects, does not depend on temperature and determines the residual resistivity at $T=0$. The temperature dependences of last two terms, $\rho_e$\ (due to electron-electron scattering) and $\rho_{sf}$\ (due to spin-fluctuation scattering), have different forms: $\rho_e(T)\propto T^2$\ \cite{Abrikosov2017}, while $\rho_{sf}(T)\propto T$\ \cite{Moriya1990,Kemper2011,Fernandes2011}. In general, the temperature dependence of resistivity has a form:
\begin{equation}
\rho(T) = \rho_o+A_1T+A_2T^2
\label{eq2}
\end{equation}
Thus, by analyzing the experimental temperature dependence of the resistivity of the sample under study, it is possible to find out which scattering mechanism is predominant. Moreover, knowing the effective mass and carrier concentration, an estimate of the total dissipation rate $\tau^{-1}$\ can be made using Equation (\ref{eq1}).

In present study the character of the temperature dependence of electrical resistivity $\rho(T)$\ was determined from the results of the $R(T)$\ measurement in the range from 4 to 300~K. Resistance measurements were carried out using a standard four-probe method at a constant current of about 1\,mA magnitude. Current and potential contacts were attached using silver conductive paste on the $ab$\ plane of the crystal. In this way, the longitudinal component of the resistance $R_{ab}$\ was measured, which was then converted into resistivity $\rho$\ taking into account the dimensions of crystal. 

To record microwave absorption (MWA), we used a standard BER-418s spectrometer from Bruker, operating at a frequency of about 9.5 GHz. The amplitude of microwave losses $A_{mwa}$\ in the spectrometer cavity with a sample was measured in the temperature range from 4.2~K to 200~K.

It is known that in a conductive material the absorption of microwaves occurs in the skin layer. Therefore, the amplitude of the MWA is proportional to the volume of the skin layer, provided that the skin-depth $\delta$\ is much greater than the electron mean free path $l_e$\ \cite{ Klein1993}. Consequently, a change in the skin-depth $\delta$\ with decreasing temperature results in a temperature dependence of the amplitude of the microwave absorption signal $A_{mwa}$. Since the skin-depth is determined by the resistivity $\rho$\ through the equation $\delta = c \sqrt{\rho \over 2\pi\omega\mu_0}$\ (here $c$\ is the speed of light, $\omega $\ is the frequency and $\mu_0$\ is the permeability of vacuum), then it turns out that $A_{mwa} \propto \sqrt{\rho}$. This refers to the contribution of ohmic losses to microwave absorption. In addition, the MWA amplitude is affected by magnetic losses, which are proportional to the static magnetic susceptibility $\chi_0$. Their contribution depends on the ratio of the measurement frequency $\omega$\ and the spin scattering rate, which is approximately equal to $\tau^{-1}_{sf}$. Taking all this into account, the microwave absorption amplitude takes the form:
\begin{equation}
    \label{eq3}
    A_{mwa}(T) = \sqrt{\frac{\mu_0\omega_0}{2}} \sqrt{\rho(T)} \left(1+\frac{1}{2} (\chi^\prime(T)-\chi^{\prime\prime}(T))\right)
\end{equation}
with real $\chi^\prime$\ and imaginary $\chi^{\prime\prime}$\ components of magnetic susceptibility, which are expressed through the formulas:
  \begin{equation}
    \label{eq4}
    \chi^{\prime}=\chi_0\frac{1}{1+(\omega\tau_{sf})^2}
\end{equation}

\begin{equation}
    \label{eq5}
	\chi^{\prime\prime}=\chi_0\frac{\omega\tau_{sf}}{1+(\omega\tau_{sf})^2} 
\end{equation}

\noindent The Equation (\ref{eq3}) was obtained by analogy with what S. Barnes did to describe the spin resonance of metal samples with magnetic impurities \cite{Barnes1981}. 

Using methods of measuring DC resistivity and microwave absorption, we studied the effect of spin fluctuations on the scattering of current carriers in the BaFe$_2$As$_2$\ crystals with different cobalt concentrations. The method of growing crystals is described in detail in the article \cite{Aswartham2011}, and their transport and magnetic properties are presented in Refs. \cite{Aswartham2011,Grafe2014}. 

\section{Results and discussion}

We will use below the temperature dependence of DC resistivity and microwave absorption, obtained for two Ba(Fe$_{1-x}$Co$_x$)$_2$As$_2$\ crystals with $x=0.05$\ (underdoped) and $x=0.075$\ (overdoped), to demonstrate their salient features and analysis. 

\begin{figure}[t]
\centering
\includegraphics[width=0.7\textwidth]{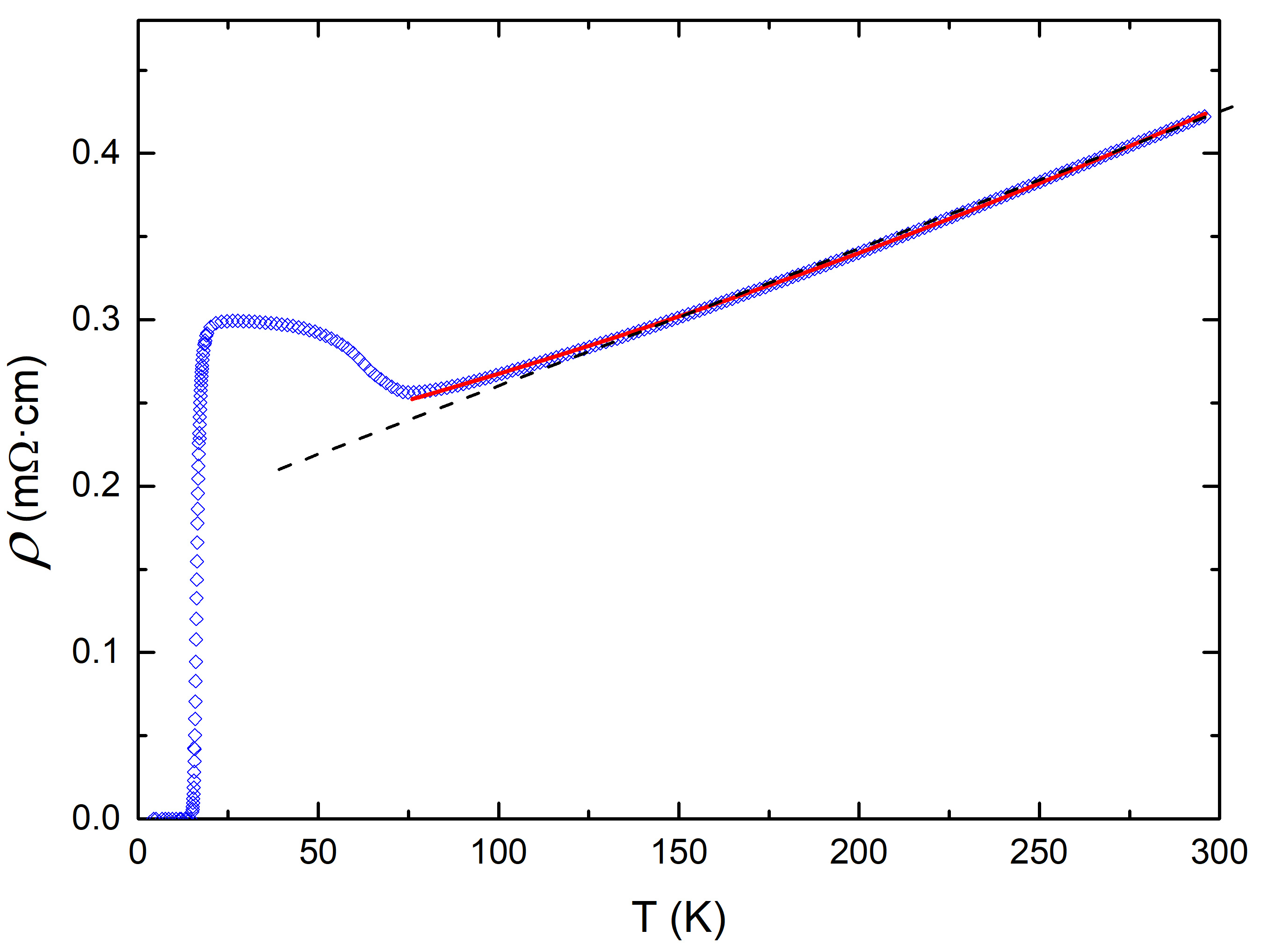}
\caption{Temperature dependence of the resistivity of the Ba(Fe$_{0.95}$Co$_{0.05}$)$_2$As$_2$\ crystal (diamonds). The dashed straight line is drawn along the experimental points obtained at high temperatures, $T>115$\,K. Red solid line is the fit of Equation \ref{eq2} to data points of temperature range from 75\,K to 300\,K.} 
\label{fig1}
\end{figure}

The resistivity versus temperature for the Ba(Fe$_{0.95}$Co$_{0.05}$)$_2$As$_2$\ sample is shown in Figure \ref{fig1}. This dependence has all the features characteristic of crystals Ba(Fe$_{1-x}$Co$_x$)$_2$As$_2$\ with a cobalt impurity concentration less than optimal, $x<0.07$ \cite{Doiron2009,Nakajima2014,Ni2008,Ning2009,Wang2009}. Namely: metallic run with a positive slope in the interval $T_s<T<300$\,K with linear dependence $\rho(T)\propto T$\ at $T>115$\,K and deviation from linearity when approaching the point of structural transition $T_s\approx75$\,K upon decreasing temperature. When $T_s$\ is reached, the resistance makes a sharp jump upward, after which it changes slightly, then sharply drops to 0 upon transition to the superconducting state at $T=T_c$. It is generally accepted that the $\rho(T)$\ dependence is described by the Equation (\ref{eq2}) and its linear part is due to carrier scatter by spin fluctuations \cite{Ni2008,Ning2009,Wang2009,Doiron2009,Kemper2011}. The presence of spin fluctuations and their effect on the transport and magnetic properties of iron arsenides has been confirmed by many experimental methods, including inelastic neutron scattering (INS) \cite{Matan2010,Lester2010,Wang2013} and nuclear magnetic resonance (NMR) \cite{Ning2009,Ning2010,Grafe2014,Dioguardi2015}. In particular, they determine the unusual form of the temperature dependence of magnetic susceptibility: $\chi_0(T)\propto T$\ instead of the Curie-Weiss law. This form of dependence is due to that the exchange interaction $J_1$\ between the magnetic moments of iron ions in the FeAs layers does not lead to long-range magnetic ordering even when the temperature drops below $\Theta$\ ($k_B\Theta\approx J_1=43$\,meV \cite{Lester2010}). It is because of the prohibition of such ordering in low-dimensional systems (the Mermin-Wagner theorem \cite{Halperin2019}), and the FeAs layers in Ba(Fe$_{1-x}$Co$_x$)$_2$As$_2$\ are quasi-2D due to the weak connection $J_{\perp}$\ between layers. At the same time, sufficiently strong antiferromagnetic correlations lead to the formation of singlet pairs and a decrease in magnetic susceptibility with lowering temperature \cite{Korshunov2009,Zhang2009,Harnagea2011}.

According to popular opinion, at low temperatures ($T<100$\,K), the electron-electron scattering mechanism becomes more efficient than scattering by spin fluctuations, and the $\rho(T)$\ dependence turns into quadratic \cite{Moriya1990,Doiron2009,Barisic2010}. The reason for this crossover is not entirely clear and leaves doubt as to its correctness. Here we propose another idea to explain the behavior of the $\rho(T)$\ dependence at low temperatures.

In our opinion, it is more natural to associate the deviation from linearity with the influence of nematic fluctuations. The fact is that the magnetic fluctuations, which determine the resistivity value, have a highly anisotropic stripe-type order parameter. Therefore, the scattering rates on them for two crystallographic directions ($a$\ and $b$) are very different: $\rho^b>\rho^a$. A theoretical description of this effect is given in Ref.\cite{Fernandes2011}. This is experimentally confirmed by measurements of the resistivity in two directions, $\rho^a$\ and $\rho^b$, of Ba(Fe$_{1-x}$Co$_x$)$_2$As$_2$\ crystals subjected to uniaxial compression (tension), which leads to their detwinning \cite{Ying2011,Ishida2013}. As a result, the dependences $\rho^a(T)$\ and $\rho^b(T)$\ are very different from each other both at $T<T_s$\ and at $T>T_s$, and the temperature at which resistivity anisotropy appears is tens of degrees higher than $T_N$\ and $T_s$. In this case, $\rho^a(T)$\  remains practically linear as the temperature decreases almost to the superconducting transition $T_c$, while $\rho^b(T)$\ deviates from linearity with the emergence of nematic fluctuations and experiences an upturn at temperature close to $T_N$. In twinned crystals, which have not undergone uniaxial compression, $\rho(T)$\ behaves as something between $\rho^a(T)$\ and $\rho^b(T)$, that is, all deviations from linearity also occur. They are not as large as $\rho^b(T)$, but they are quite comparable to it in the degree of nematic fluctuation effect. Thus, the behavior of $\rho(T)$\ can be well explained by the influence of nematic (anisotropic magnetic) fluctuations, without using the assumption of a transition to the Fermi-liquid regime. 

\begin{figure}[t]
\centering
\includegraphics[width=0.7\textwidth]{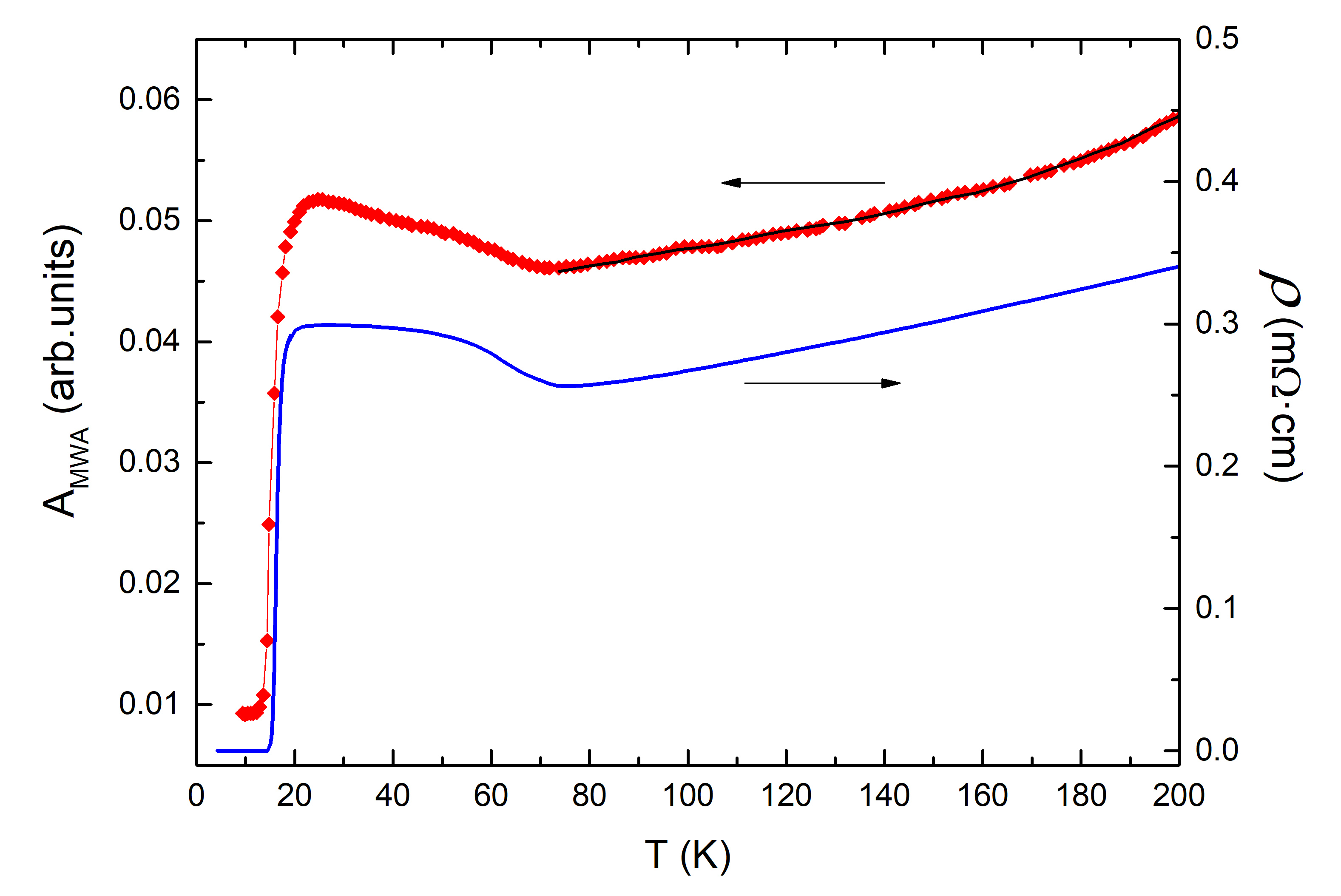}
\caption{Temperature dependence of the MWA amplitude (red diamonds) and resistivity (blue curve) of the Ba(Fe$_{0.95}$Co$_{0.05}$)$_2$As$_2$\ crystal. Black solid line is the fit of Equation \ref{eq3} to the MWA data points of temperature range from 70\,K to 200\,K.} 
\label{fig2}
\end{figure}

As mentioned above, microwave absorption data provide additional information about fluctuations. The temperature dependence of the MWA amplitude for the Ba(Fe$_{0.95}$Co$_{0.05}$)$_2$As$_2$\ sample is presented in Figure \ref{fig2} along with $\rho(T)$. To analyze the $A_{MWA}(T)$\ behavior, Equation (3) with definitions (4) and (5) is used. In so doing, we assume a linear dependence of $\rho(T)$\ and $\chi_0(T)$. The temperature dependence of the scattering rate is used in the form proposed in Ref.\cite{Doiron2009}: 
\begin{equation}
  \label{eq6}
	\tau^{-1}(T)=\frac{a\cdot T}{(T+\Theta)^{1/2}}+b\cdot T^2  ,
\end{equation}
where $\Theta\approx {J_1\over k_B} \approx 500$\,K is the mean-field temperature of magnetic ordering in the $ab$\ plane that did not occur due to the reasons noted above (value of $J_1$\ is taken from Ref.\cite{Lester2010}). The first term on the right side of the Equation (\ref{eq6}) is related to scattering by spin fluctuations, and the second one describes electron-electron scattering. The result of fitting is shown in Figure \ref{fig2} with a solid black curve. Note that when calculating without restricting the parameters, the value of the parameter $b$\ is always less than $a$\ by several orders of magnitude, and if we take $b = 0$, then the calculated curve virtually does not change. This indicates that there is no need to involve the Fermi-liquid scattering mechanism to describe ohmic losses and microwave absorption in iron pnictides, despite the fact that the $A_{MWA}(T)$\ dependence is nonlinear. The magnitude of $\tau^{-1}$\ will be discussed below. 

\begin{figure}[t]
\centering
\includegraphics[width=0.7\textwidth]{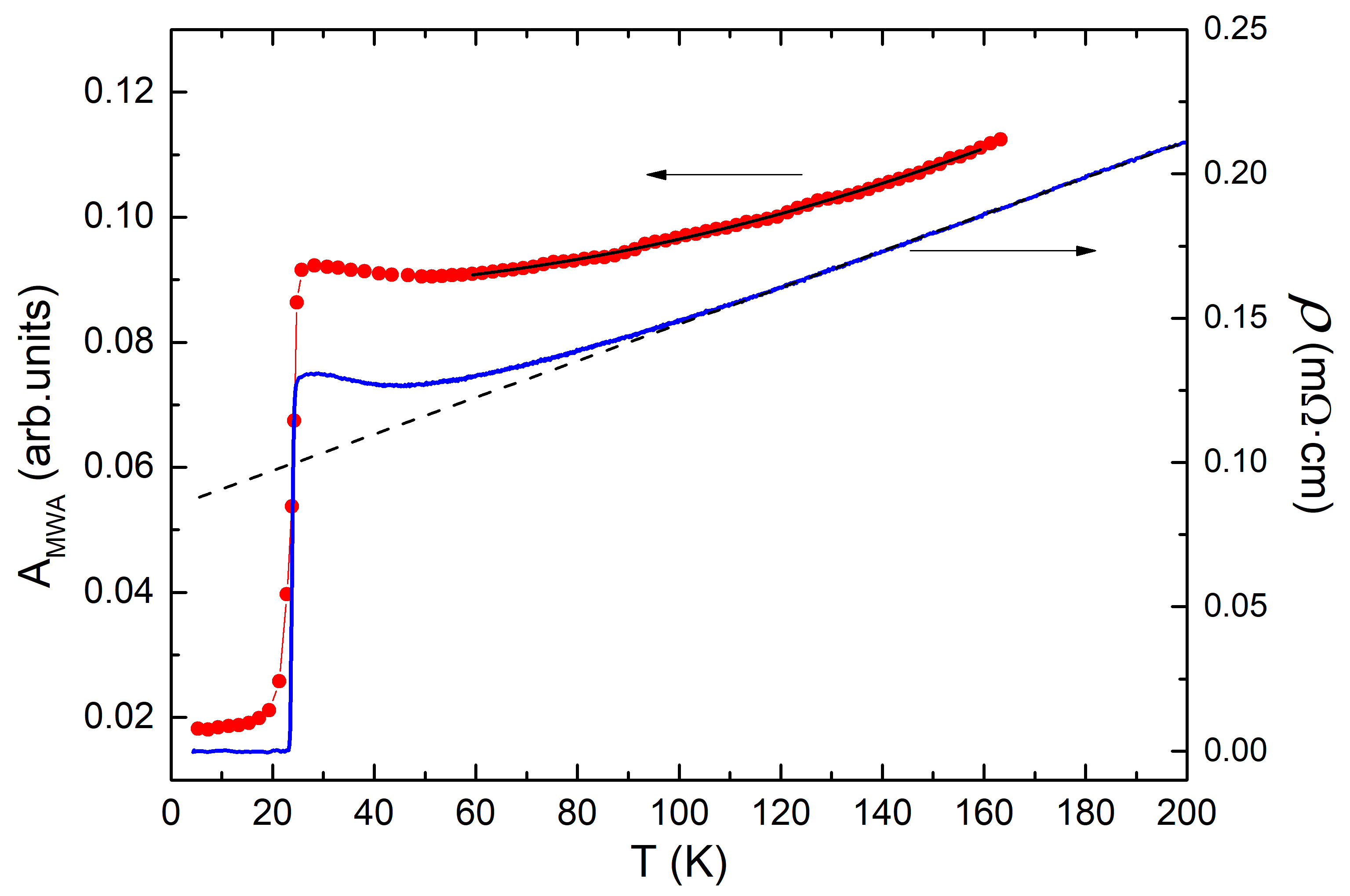}
\caption{Temperature dependence of resistivity (blue solid line) and the MWA amplitude (red circles) of the Ba(Fe$_{0.925}$Co$_{0.075}$)$_2$As$_2$\ crystal. Black solid line is the fit of Equation \ref{eq3} to the MWA data in the temperature range from 60\,K to 165\,K. A dashed straight line is drawn to show a linear section of $\rho(T)$.} 
\label{fig3}
\end{figure}

 The change of temperature dependence of resistivity and microwave absorption in a crystal Ba(Fe$_{1-x}$Co$_x$)$_2$As$_2$\ when the cobalt concentration increases to values higher than optimal ($x>0.07$) will be considered below using the example of the Ba(Fe$_{0.925}$Co$_{0.075}$)$_2$As$_2$\ sample. In accordance with the phase diagram \cite{Ni2008,Aswartham2011}, neither a structural transition nor magnetic ordering occurs in a sample of such composition. The dependences $\rho(T)$\ and $A_{MWA}(T)$\ obtained for this sample, presented in Figure 3, are well consistent with this statement. There are no features on both curves due to changes in structure or magnetic state. There is only a sharp drop in value (to 0 in the case of resistivity) when the sample transits to the superconducting state, $T_c=24$\,K. $\rho(T)$\ has a fairly long linear section, $100\div300$\,K, and upturn at $T=43$\,K. $A_{MWA}(T)$\ is nonlinear over the entire measurement temperature range, from $T_c$\ up to 165\,K. The calculation of this dependence using Equation (\ref{eq3}) and the definitions of its parameters by formulas (4-6) is shown by black curve in the Figure \ref{fig3}. 

One of the most important parameters of the studied compounds, which determines the value and behavior of both resistivity (Eq.\ref{eq1}) and microwave absorption (Eq.\ref{eq3}), is the scattering rate $\tau^{-1}$\ (or scattering time $\tau$). The parameter values obtained by fitting the theoretical dependence $A_{MWA}(T)$\ (Eq.(\ref{eq3})) to the experimental data were as follows: $\tau^{-1}$\ changes from $2.1\cdot10^{10}$\,s$^{-1}$\ to $6.3\cdot10^{10}$\,s$^{-1}$\ with increasing temperature in the range of 75\,K - 185\,K in case of sample Ba(Fe$_{0.95}$Co$_{0.05}$)$_2$As$_2$. And it varies from $2\cdot10^{10}$\,s$^{-1}$\ to $1.5\cdot10^{11}$\,s$^{-1}$\ at $T=60\div 185$\,K for the Ba(Fe$_{0.925}$Co$_{0.075}$)$_2$As$_2$\ sample. The obtained values of scattering rates are significantly lower than their estimates made for similar samples based on DC transport measurements and presented in literature: $3\cdot10^{13}$\,s$^{-1}$\ at 300\,K  \cite{Barisic2010} and $3.5\cdot10^{12}$\,s$^{-1}$\ at 100\,K \cite{Rullier2009}. This discrepancy can be understood if we take into account that estimates from transport measurement data give the total dissipation rate $\tau^{-1}$, which is determined by all mechanisms, including an impurity one. And the values obtained from the MWA fitting relate only to scattering by spin fluctuations, $\tau_{sf}^{-1}$. This explanation is consistent with the results of ultrafast time-resolved polarimetry \cite{Patz2014}, which revealed a two-step recovery of spin system of Ba(Fe$_{1-x}$Co$_x$)$_2$As$_2$\ excited by a femtosecond pulse of polarized light. At the first step, rapid thermalization of charges occurs within $10^{-13}$\,s after the exciting pulse. The second step is a slow recovery via nematic fluctuations over a period of $10^{-11}$\,-\,$10^{-10}$\,s. It can be seen that the $\tau^{-1}_{sf}$\ values obtained from the $A_{MWA}(T)$\ analysis correspond to the second stage of relaxation of the spin system. 

\section{Conclusion}

In order to clarify what mechanisms of current carrier scattering determine the magnitude and behavior of resistance and microwave absorption in crystals Ba(Fe$_{1-x}$Co$_x$)$_2$As$_2$, we analyzed the temperature dependences $\rho(T)$\ and $A_{MWA}(T)$\ of samples with cobalt concentration $x=0.05$\ and 0.075. The results of analysis allow us to conclude that the deviation of the $\rho(T)$\ dependence from the linear behavior at $T\leq 100$\,K is not related to the transition to the Fermi-liquid regime with $\rho(T)\propto T^2$, but is due to the emerging nematic fluctuations. The rate of scattering by spin fluctuations $\tau^{-1}_{sf}$, determined by describing the temperature dependence of the MWA amplitude, indicates that they take nematic character as the temperature decreases down to the SDW ordering point.

\bigskip
\bibliographystyle{apsrev4-2}
\bibliography{art4pns}

\end{document}